\documentclass[superscriptaddress,twocolumn]{revtex4}
\usepackage{bm}
\usepackage{graphicx}
\usepackage{hyperref}

\begin{document}

\title{Comment on ``HADOKEN: An open-source software package for predicting electron confinement effects in various nanowire geometries and configurations''}

\author{I.A. Kokurin}
\email[E-mail:]{kokurinia@mail.ru} \affiliation{Institute of Physics
and Chemistry, Mordovia State University, 430005 Saransk, Russia}
\affiliation{Ioffe Institute, 194021 St. Petersburg, Russia}
%\affiliation{St. Petersburg Electrotechnical University ``LETI'',
%197376 St. Petersburg, Russia}

\begin{abstract}
In a recent work [C. Chevalier, B. M. Wong, Comput. Phys. Commun.
{\bf 274}, 108299 (2022); arXiv: 2203.05233 [cond-mat.mes-hall]] the
interesting and popular problem was considered. Authors attempted to
solve the self-consistent Schr\"{o}dinger-Poisson problem for an
effective mass electron in a core-shell semiconductor nanowire. The
corresponding MATLAB-based software package was presented. However,
an incorrect solution of the Schr\"{o}dinger equation invalidates
the whole result. Here we point out the corresponding error and
possible ways to fix it.
\end{abstract}

\date{\today}

\maketitle

In recent years, many calculations of the spectrum of charge
carriers and wave functions in semiconductor nanowires (NWs) have
been
made~\cite{Royo2014,Degtyarev2017,Wojcik2018,Sitek2018,Kokurin2020},
including those in core-shell structures. In this context, an
emergence of a tool that allows one to calculate the electronic
states in the core-shell NWs is very useful and timely. Such a tool
appeared quite recently~\cite{Chevalier2022}. This free software is
based on MATLAB and is called HADOKEN (High-level Algorithms to
Design, Optimize, and Keep Electrons in Nanowires). However, the
solution of the spectral problem in Ref.~\onlinecite{Chevalier2022}
was incorrect. This led to the incorrectness of the whole
self-consistent Schr\"{o}dinger-Poisson procedure. In this Comment
we point out the corresponding error and possible ways to fix it.

The effective mass Hamiltonian for an electron of simple isotropic
conduction band in a core-shell NW is given by

\begin{equation}
H=-\frac{\hbar^2}{2}\nabla\frac{1}{m^*(x,y)}\nabla+V_T(x,y),
\end{equation}
where $m^*(x,y)$ is the coordinate-dependent effective mass, and
$V_T(x,y)$ is the total potential, including the quantum confinement
due to the conduction band profile in a core-shell heterostructure
and the electrostatic potential. Since the translational invariance
in the NW-axis direction (the $z$-axis) is conserved, we can write
the envelope wave function in the following form

\begin{equation}
\Psi_{nk}(x,y,z)=\frac{1}{\sqrt L}e^{ikz}\psi_n(x,y).
\end{equation}

Here $\hbar k$ is the longitudinal momentum and $n$ enumerates the
one-dimensional subbands.

Substituting this wave function into the Schr\"{o}dinger equation,
$H\Psi=E\Psi$, we find

\begin{eqnarray}
\nonumber &-&\frac{\hbar^2}{2}\left[\frac{\partial}{\partial
x}\frac{1}{m^*(x,y)}\frac{\partial\psi_n(x,y)}{\partial
x}+\frac{\partial}{\partial
y}\frac{1}{m^*(x,y)}\frac{\partial\psi_n(x,y)}{\partial
y}\right]\\
&+&\left(V_T(x,y)+\frac{\hbar^2k^2}{2m^*(x,y)}\right)\psi_n(x,y)=E\psi_n(x,y).
\end{eqnarray}

One can see, that, due to the position-dependent effective mass, the
variables are not separated. In order to find the subband energy
spectrum one have to solve the problem numerically for each fixed
$k$ value. The energy eigenvalue is $E_n(k)$, and the corresponding
eigenfunction can be rewritten as $\psi_n(x,y;k)$, which indicates a
parametric dependence on $k$. A possible solution of this problem
for an electron in core-multi-shell NW structure was discussed in
Ref.~\onlinecite{Rudakov2019}.

However, the authors of Ref.~\onlinecite{Chevalier2022} wrote a very
strange answer. They formally separated longitudinal and transverse
motion, and wrote: $E_n(k)=E_n+E_z$, where $E_n$ is the $n$-th
subband bottom energy and $E_z=\frac{\hbar^2k^2}{2m^*(x,y)}$. The
dependence of the eigenvalue on the spatial variables is a very
strange answer!!!

Thus, the above two things [(i) an incorrectly found energy spectrum
and (ii) an unaccounted for parametric dependence of the wave
function on the longitudinal momentum] lead to a wrong result,
starting with Eq.~(14) of Ref.~\onlinecite{Chevalier2022}.

In the case of a step-like effective mass profile, when $m_c$ and
$m_s$ are the effective masses of the core and shell material,
respectively, the energy spectrum can be found in the form

\begin{equation}
E_n(k)=E_n+\frac{\hbar^2k^2}{2m^*_n},
\end{equation}
where $E_n$ is again the subband bottom energy, and $m^*_n$ is the
$n$-th subband effective mass, which, due to the penetration of the
wave function into the barrier regions, is something between $m^*_c$
and $m^*_s$.

Taking into account the above corrections, Eq.~(15) of
Ref.~\onlinecite{Chevalier2022} should be replaced by (here the
notation of Ref.~\onlinecite{Chevalier2022} is used)

\begin{equation}
n_e(\widetilde{x},\widetilde{y})=\frac{1}{\pi l_0^2}\sum_n\int dk_z
|\psi_n(\widetilde{x},\widetilde{y};k_z)|^2f(E_n(k_z),\mu,T),
\end{equation}
where $f(E,\mu,T)$ is the Fermi distribution function, $\mu$ and $T$
are the chemical potential and temperature, respectively. Such
corrections, apparently, complicate the numerical calculation, but
make it correct.

I guess that appropriate changes should be made in the code of the
HADOKEN software. Moreover, the Computer Physics Communications and
its library allow to improve the software code.

%\bibliography{Kokurin}

\begin{thebibliography}{7}
\expandafter\ifx\csname
natexlab\endcsname\relax\def\natexlab#1{#1}\fi
\expandafter\ifx\csname bibnamefont\endcsname\relax
  \def\bibnamefont#1{#1}\fi
\expandafter\ifx\csname bibfnamefont\endcsname\relax
  \def\bibfnamefont#1{#1}\fi
\expandafter\ifx\csname citenamefont\endcsname\relax
  \def\citenamefont#1{#1}\fi
\expandafter\ifx\csname url\endcsname\relax
  \def\url#1{\texttt{#1}}\fi
\expandafter\ifx\csname urlprefix\endcsname\relax\def\urlprefix{URL
}\fi \providecommand{\bibinfo}[2]{#2}
\providecommand{\eprint}[2][]{\url{#2}}

\bibitem[{\citenamefont{Royo et~al.}(2014)\citenamefont{Royo, Bertoni, and
  Goldoni}}]{Royo2014}
\bibinfo{author}{\bibfnamefont{M.}~\bibnamefont{Royo}},
  \bibinfo{author}{\bibfnamefont{A.}~\bibnamefont{Bertoni}}, \bibnamefont{and}
  \bibinfo{author}{\bibfnamefont{G.}~\bibnamefont{Goldoni}},
  \bibinfo{journal}{Phys. Rev. B} \textbf{\bibinfo{volume}{89}},
  \bibinfo{pages}{155416} (\bibinfo{year}{2014}).

\bibitem[{\citenamefont{Degtyarev et~al.}(2017)\citenamefont{Degtyarev,
  Khazanova, and Demarina}}]{Degtyarev2017}
\bibinfo{author}{\bibfnamefont{V.~E.} \bibnamefont{Degtyarev}},
  \bibinfo{author}{\bibfnamefont{S.~V.} \bibnamefont{Khazanova}},
  \bibnamefont{and} \bibinfo{author}{\bibfnamefont{N.~V.}
  \bibnamefont{Demarina}}, \bibinfo{journal}{Scientific Reports}
  \textbf{\bibinfo{volume}{7}}, \bibinfo{pages}{3411} (\bibinfo{year}{2017}).

\bibitem[{\citenamefont{W\'ojcik et~al.}(2018)\citenamefont{W\'ojcik, Bertoni,
  and Goldoni}}]{Wojcik2018}
\bibinfo{author}{\bibfnamefont{P.}~\bibnamefont{W\'ojcik}},
  \bibinfo{author}{\bibfnamefont{A.}~\bibnamefont{Bertoni}}, \bibnamefont{and}
  \bibinfo{author}{\bibfnamefont{G.}~\bibnamefont{Goldoni}},
  \bibinfo{journal}{Phys. Rev. B} \textbf{\bibinfo{volume}{97}},
  \bibinfo{pages}{165401} (\bibinfo{year}{2018}).

\bibitem[{\citenamefont{Sitek et~al.}(2018)\citenamefont{Sitek,
  Urbaneja~Torres, Torfason, Gudmundsson, Bertoni, and Manolescu}}]{Sitek2018}
\bibinfo{author}{\bibfnamefont{A.}~\bibnamefont{Sitek}},
  \bibinfo{author}{\bibfnamefont{M.}~\bibnamefont{Urbaneja~Torres}},
  \bibinfo{author}{\bibfnamefont{K.}~\bibnamefont{Torfason}},
  \bibinfo{author}{\bibfnamefont{V.}~\bibnamefont{Gudmundsson}},
  \bibinfo{author}{\bibfnamefont{A.}~\bibnamefont{Bertoni}}, \bibnamefont{and}
  \bibinfo{author}{\bibfnamefont{A.}~\bibnamefont{Manolescu}},
  \bibinfo{journal}{Nano Letters} \textbf{\bibinfo{volume}{18}},
  \bibinfo{pages}{2581} (\bibinfo{year}{2018}).

\bibitem[{\citenamefont{Kokurin}(2020)}]{Kokurin2020}
\bibinfo{author}{\bibfnamefont{I.~A.} \bibnamefont{Kokurin}},
  \bibinfo{journal}{Semiconductors} \textbf{\bibinfo{volume}{54}},
  \bibinfo{pages}{1897} (\bibinfo{year}{2020}).

\bibitem[{\citenamefont{Chevalier and Wong}(2022)}]{Chevalier2022}
\bibinfo{author}{\bibfnamefont{C.}~\bibnamefont{Chevalier}} \bibnamefont{and}
  \bibinfo{author}{\bibfnamefont{B.~M.} \bibnamefont{Wong}},
  \bibinfo{journal}{Comput. Phys. Commun.} \textbf{\bibinfo{volume}{274}},
  \bibinfo{pages}{108299} (\bibinfo{year}{2022}).

\bibitem[{\citenamefont{Rudakov and Kokurin}(2019)}]{Rudakov2019}
\bibinfo{author}{\bibfnamefont{A.~O.} \bibnamefont{Rudakov}} \bibnamefont{and}
  \bibinfo{author}{\bibfnamefont{I.~A.} \bibnamefont{Kokurin}},
  \bibinfo{journal}{Semiconductors} \textbf{\bibinfo{volume}{53}},
  \bibinfo{pages}{2137} (\bibinfo{year}{2019}), \bibinfo{note}{the more correct
  version is arXiv:2101.06532 [cond-mat.mes-hall].}

\end{thebibliography}

\end{document}